\documentclass[a4paper]{jpconf}
\usepackage{graphicx}
\usepackage{iopams}

\begin{document}

\title{Confronting LHC data with the statistical hadronization model}

\author{J Stachel$^1$, A Andronic$^2$, P Braun-Munzinger$^{2,3,4}$, K
  Redlich$^5$} 
\address{$^1$Physikalisches Institut, Universit\"at
  Heidelberg, Heidelberg, Germany} 
\address{$^2$Research Division and
  ExtreMe Matter Institute EMMI, GSI Helmholtzzentrum f\"ur
  Schwerionenforschung, Darmstadt, Germany} 
\address{$^3$ Technische
  Universit\"at Darmstadt, Darmstadt, Germany}
\address{$^4$ Frankfurt Institute for Advanced Studies, J.W. Goethe
  Universit\"at, Frankfurt, Germany}
\address{$^5$ Institute
  of Theoretical Physics, University of Wroclaw, Wroclaw, Poland}

\ead{stachel@physi.uni-heidelberg.de}

\begin{abstract}

The most recent data from the CERN LHC are compared with calculations
within the statistical hadronization model. The parameters temperature
und baryon chemical potential are fitted to the data. The best fit
yields a temperature of 156 MeV, slightly below the expectation from
RHIC data. Proton yields are nearly three standard deviations below
this fit and possible reasons are discussed.
 
\end{abstract}

\section{Introduction}

In ultra-relativistic nuclear collisions the phase diagram of strongly
interacting matter is studied \cite{pbmwambach}. One of the
interesting questions is about the location of the boundary between
hadronic matter and the quark-gluon plasma (QGP). Over the past 20
years the understanding has arisen, that hadron yields can be very
successfully compared to a simple statistical model. For central
collisions of heavy nuclei (Au or Pb) and energies $E_{lab}/A \geq$ 5
GeV, the grand canonical ensemble gives an appropriate representation.
The resulting parameters, temperature $T$ and baryon chemical
potential $\mu_b$, represent the so-called chemical freeze-out. It was
further realized \cite{rapidtherm} that the very high densities close
to the transition temperature, when a QGP hadronizes, and the rapid
fall-off of density below the transition temperature lead to rapid
equilibration and chemical freeze-out only a few MeV below the
transition temperature. The current understanding of the connection
between the QCD phase diagram and chemical freeze-out is summarized in
\cite{gerry85}.

The statistical hadronization model was first successfully applied to
data from ultra-relativistic nuclear collisions from the AGS
\cite{thermags} and not much later to first data from the CERN SPS
\cite{thermsps}. Both comparisons yielded excellent agreement. Later,
a much more complete set of hadron yields was obtained from the SPS
experiments, including data from several energies below the maximum
energy down to $E_{lab}/A = $ 20 GeV. And from the year 2000 on an
increasing set of hadron yields from Au+Au collisions at RHIC became
available and was compared to statistical model calculations
\cite{thermrhic}. All data sofar were found to be in very good
agreement with the grand canonical statistical model. For a summary
see \cite{thermcomp}.

Inspecting the extracted statistical model parameters as a function of
center of mass energy, two characteristic features arose: (i) The
baryon chemical potential drops monotonically with increasing energy
and (ii) the temperature initially increases but appears to level off
at center of mass energies per colliding nucleon pair of about 20 GeV
at a temperature around 160 MeV. This led to a prediction for LHC
\cite{lhcpred} that chemical freeze-out would occur at $\mu_b =
0.8^{+1.2}_{-0.6}$ MeV and $T = 161 \pm$ 4 MeV.

\section{LHC data}

\begin{figure}[hbt]
\begin{center}
\includegraphics[width=13cm]{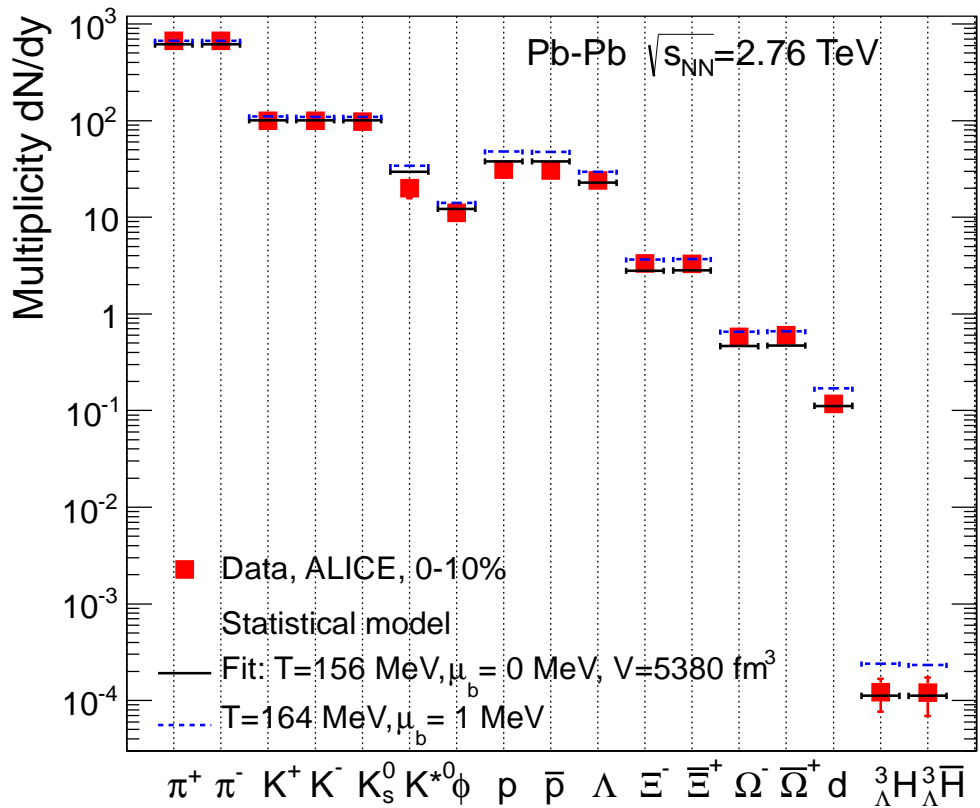}
\end{center}
\caption{\label{thermalfitlhc} Hadron yields from ALICE at the LHC
  \cite{pikp,k0lambda,multistrange,k*phi,d,nuclei} and fit with the
  statistical hadronization model. In addition to the fit, yielding
  $T$=156 MeV, also results of the model for $T$ = 164 MeV are shown,
  normalized to the value for $\pi^+$. The data point for the K$^{0*}$
  is not included in the fit.}
\end{figure}

\begin{figure}[hbt]
\begin{center}
\includegraphics[width=13cm]{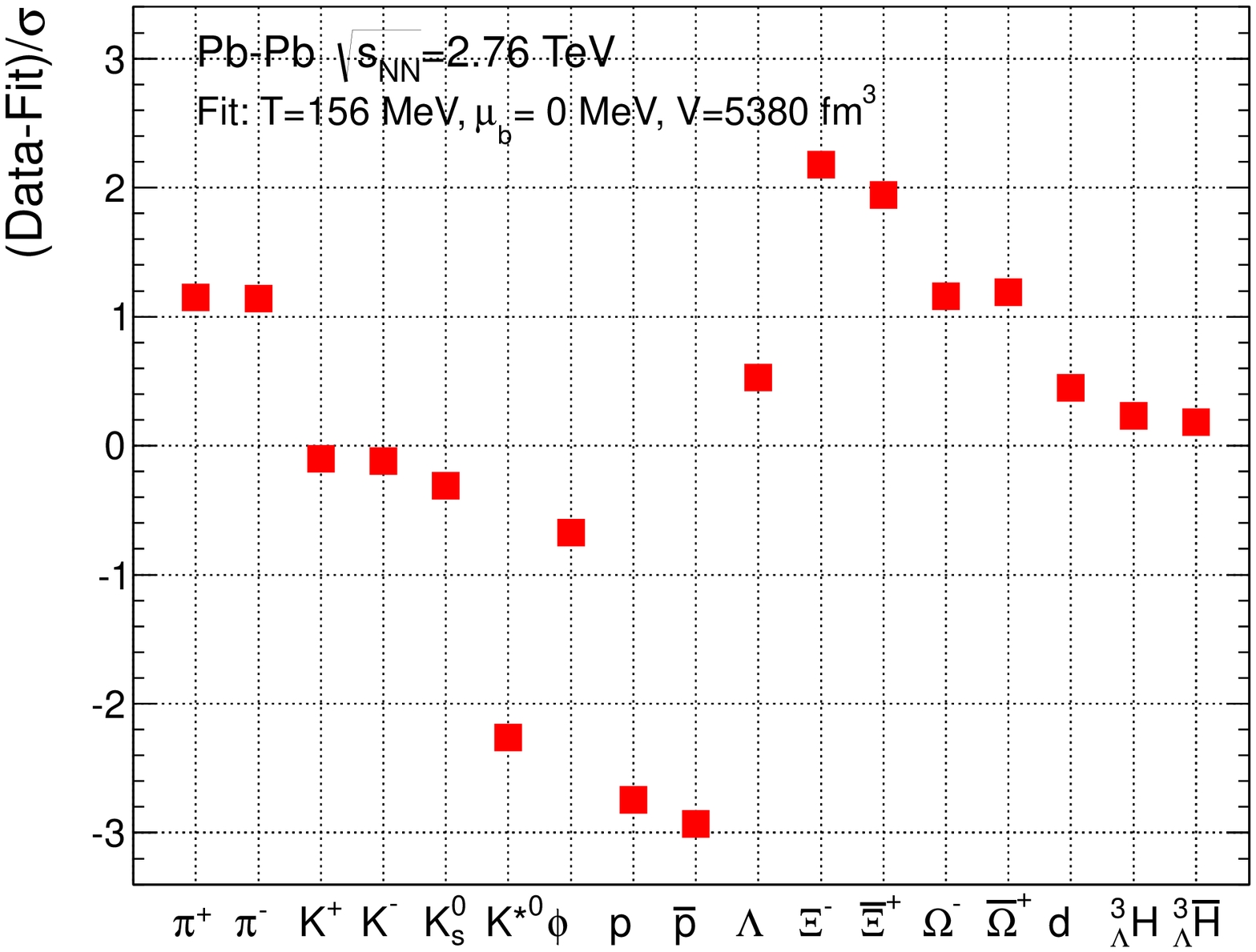}
\end{center}
\caption{\label{sthermalfitlhc} Deviations between thermal fit and
  data normalized to the error of the data points.}
\end{figure}

First results of ALICE at the LHC were discussed in the framework of
the statistical hadronization model in \cite{qm2011,pbmqm2012}. Now a
much extended set of published data is available as well as some
preliminary data on the production yields of nuclei. A fit of the
statistical hadronization model to the data currently available from
ALICE at the LHC is shown in Fig.~\ref{thermalfitlhc}. The data point
for K$^{0*}$ is not included in the fit. The fit yields a baryon
chemical potential of zero and a temperature of 156 MeV with a reduced
$\chi^2$ of 2.4. A fit to the preliminary data \cite{pbmqm2012} was
significantly worse with $T$ = 152 MeV and a reduced $\chi^2 =
4.3$. In Fig.~\ref{sthermalfitlhc} the deviations between fit and data
are shown. The proton and antiproton yields are under the model by 18.0
and 19.4~\% which, due to the small experimental errors, amounts to a
deviation of 2.7 and 2.9 sigma, respectively. The cascade yields, on
the other hand, are above the model by about 2 sigma. Otherwise the
agreement of data and fit is excellent. The deviation for the K$^{0*}$
meson should be ignored; as a strongly decaying resonance it's yield
can be significantly modified after chemical freeze-out.

To demonstrate the sensitivity of the model prediction to the
temperature, we also show in Fig.~\ref{thermalfitlhc} results for a
statistical model calculation using $T$ = 164 MeV and $\mu_b$ = 1
MeV. This higher temperature increases the disagreement for the
antiprotons to about 50 \% and the yields of nuclei are much
overpredicted. Entirely leaving protons out of the fit, the
temperature would increase by 2 MeV to $T$ = 158 MeV with an otherwise
prefect fit with a reduced $\chi^2$ less than one. In that sense one
could talk about a proton anomaly, albeit not a very strong one (see
above).

\section{Comments on the proton anomaly}

Already for the RHIC data there is some indication of a low proton
yields as compared to the statistical model (see
e.g. \cite{pbmqm2012}). But due to bigger uncertainties in removing
the contributions from feeding by weak decays - the published hadron
yields were obtained without vertex detectors - there are deviations
between experiments and no clear picture emerges. 

In the following, a few arguments in connection with the close to 3
standard deviation difference of the experimental proton yields from
the statistical hadronization model calculations will be made.

The hadron spectrum that enters in the statistical model incorporates
all known hadronic states contained in the 2008 compilation by the
particle data group \cite{pdg2008}. It is clear that this spectrum is
still incomplete, there are as of yet undiscovered states. A study was
made of the effect of an incomplete hadron spectrum on the K/$\pi$
ratio \cite{hagedorn_cont}. It was based on a Hagedorn resonance gas
assumption for states above 3 GeV in mass. This was found to reduce
the calculated K/$\pi$ ratio by about 15 \%. While a similar study for the
p/$\pi$ ratio does not yet exist, we would like to argue that it will
also reduce the calculated ratio; high mass hadrons will contribute at
most one (anti-)proton but multiple pions. In addition, results from
lattice QCD predict \cite{lattice_baryons} that numerous additional
baryon resonances exist at low masses, partly with high spin and
therefore degeneracy. Some of these states have already been found
\cite{thoma}. The effect of an incomplete hadronic spectrum in the
statistical hadronization model is to produce relatively too many
protons as compared to pions.

Alternatively, it has been argued, that annihilation in the hadronic
phase would reduce the number of (anti-)protons \cite{bass,
  becattini1, becattini2}. Employing UrQMD to model a hadron gas after
hadronization significant reductions in the yields of (anti-)protons
and cascades are observed for the current LHC energy. This leads to a
good description of the observed (anti-)proton yields. However,
employing this mechanism, the discrepancy for (anti-)cascades is
increased. Also, it has been noted by the authors themselves
\cite{becattini1}, that in UrQMD detailed balance is not implemented
for some of the important annihilation reactions. Already in
\cite{shuryak} it was argued, that implementing detailed balance would
not lead to a depletion of the antiprotons. The effect of annihilation
alone and of then in addition including the back reactions with full
detailed balance was studied for full SPS energy \cite{cassing} (and
also AGS energy. There it was shown that the annihilation plus back
reaction nearly fully compensate for central collisions reaching the
equilibrium value for (anti-)proton yields. In a more recent study for
collider energies it was shown \cite{pratt} that properly taking into
account the back reactions reduces the effect of annihilation in the
hadronic phase to about one half. Here, \mbox{(anti-)}protons,
lambdas, cascades and omegas are equally affected, making the
agreement for the last 2 species worse. Another argument why one
should not put too much trust in the quantitative changes of hadron
yields in the hadronic phase within the UrQMD model is the lifetime of
the fireball. From 2-pion Hanbury Brown-Twiss correlations an overall
lifetime of the system including QGP phase and hadronization of 10
fm/c is deduced \cite{alice_hbt} for central PbPb collisions at the
LHC. Coupling UrQMD to a hydrodynamics evolution the system, the
integral time until thermal freeze-out is significantly longer.

Annihilation in the hadronic phase should affect nuclei as well and it
can be seen from Fig.~\ref{sthermalfitlhc} that they are perfectly
reproduced without annihilation. One could ask, why (lightly) bound
nuclei should also follow the statistical hadronization approach. This
is plausible since hadronic reinteractions do not change the entropy
per baryon. For much lower energies this argument was already made by
Siemens and Kapusta \cite{siemens} to deduce the entropy per baryon
from the yields of light nuclei relative to protons (essentially the
d/p ratio). It has been shown \cite{coales} that for a system in
equilibrium the statistical hadronization and the coalescence approaches
agree over many orders of magnitude. In fact, the statistical
hadronization approach reproduces very well measured yields of nuclei
for central collisions at all energies from AGS up to LHC.

A puzzle is currently the centrality dependence of the proton to pion
ratio, which is increasing with centrality for the PHENIX data from
RHIC and is decreasing for the ALICE data at the LHC \cite {pikp}. This opposite
trend does not support the annihilation picture of protons in the
hadronic phase. If it is a real effect (the current significance is
only at the 2 sigma level) it has currently no physics explanation.

A proposal has been made \cite{rafelski1,rafelski} to extend the
statistical hadronization model to include out-of-equilibrium features
and according new parameters. It is therefore not surprizing that the
proton yield can be brought into agreement with such a model
calculation. A stringent test for this approach will be the yields of
light (anti-)nuclei, since then no additional free parameter is
available and they are sensitive to increasing powers of the quark
chemical potential. Already the yields of (anti-)hypertriton
presented here are in good agreement with the standard statistical
hadronization picture employed in this contribution, while they are
overpredicted by a factor of 6 in the out-of-equilibrium model of
\cite{rafelski}.

\section{Conclusion}

The ALICE data from the LHC pose with their small errors an
unprecedented test to the statistical hadronization model. An increasing
number of final data has become available over the past
year. Excellent agreement with the statistical hadronization model has
been achieved with exception of the (anti-)proton yields. Sofar no
convincing explanation for this 2.8 sigma deviation is
available. Beyond further phenomenological studies, we are looking
forward to more data and in particular also to high precision data for
the full LHC energy to become available in the next few years.

\ack

Support by the BMBF Verbundforschung and the Helmholtz Alliance EMMI
is gratefully acknowledged.

\end{document}